\documentclass[openacc]{rstransa}

\titlehead{Research}

\usepackage{comment}
\usepackage{siunitx}
\usepackage{url}
\DeclareSIUnit{\rpm}{rpm}
\DeclareSIUnit{\rad}{rad}
\usepackage{dcolumn}
\usepackage{bm}
\usepackage{soul}
\usepackage{ulem} 
\usepackage{lipsum}  
\usepackage[numbers]{natbib} 

\begin{document}

\title{Unjamming in a 3D Granular System: The Micromechanical Role of Friction in Force Distributions and Rheological Properties}
\author{
V. Salinas$^{1}$, H. Alarcón$^{2}$, E. Rojas$^{3}$, P. Gutiérrez$^{1}$ and G. Castillo$^{1}$}

\address{$^{1}$Instituto de Ciencias de la Ingeniería, Universidad de O'Higgins,  Rancagua, Chile\\
$^{2}$Departamento de Física y Química, Facultad de Ingeniería, Universidad Autónoma de Chile, Santiago, Chile\\
$^{3}$Departamento de Ingeniería Mecánica, Universidad de Antofagasta, Antogafasta, Chile}

\subject{Fluid Mechanics, Mechanics, Computer Science}


\corres{Gustavo Castillo\\
\email{gustavo.castillo@uoh.cl}}

\begin{abstract}
In this work, we investigate the unjamming transition in a three-dimensional granular system composed of frictional spheres, in which the packing fraction is systematically reduced by random particle extractions. Using Discrete Element Method (DEM) simulations, we analyze the evolution of key micro-mechanical quantities, such as the interparticle forces, the coordination number and the overall packing density as a function of the interparticle friction coefficient. Our results reveal friction-dependent relationships on structural as well as mechanical observables, and 
exhibit trends that are qualitatively consistent with observations reported in dense granular systems. These trends persist despite the very different driving mechanism considered here.

This paper is part of the thematic issue \emph{``Sand, silos and asteroids: clustering challenges in granular materials research''}.
\end{abstract}

\begin{fmtext}
\section{\label{sec:intro}Introduction}
Granular matter is a collection of macroscopic elements obeying seemingly simple interaction laws at the individual level. However, it exhibits high complexity at a large scale, partly because of phase change~\cite{Jaeger1996, Castillo2012}. Indeed, transitions happening during avalanches~\cite{daerr1999two-414,rojas2023stability-7ff,gans2023collapse-314} and collapse sinkholes  abruptly change the landscape, putting human-made infrastructures at risk~\cite{gutirrez2016oxford-a4b}. In this case, an unjamming transition occurs: a solid-like state, mechanically stable against small stresses, transitions to a liquid-like state where contacts no longer support forces~\cite{Liu1998}. In this article, we explore how the unjamming transition takes place in the presence of gravity.
\end{fmtext}

\maketitle
Systems that get stuck at low temperature (or energy), high density, and low drive can be unified as undergoing a jamming transition~\cite{Liu1998, Biroli2007}, whether they are grains, foams, emulsions, glassy molecular systems, or colloids. While jamming happens in a variety of conditions, such as shearing~\cite{OHern2003,jop2006constitutive-4b6} or gravity-free decompression of spheres~\cite{OHern2002,silbert2010jamming-20a}, unjamming under gravity has also been investigated in several configurations, including inclined planes and other gravity-driven setups~\cite{Silbert2002, Srivastava2022, lambert2025combined-910}. In our system, gravity is explicitly present and introduces a preferred direction, resulting in an anisotropic stress distribution at the macroscopic scale. In this scenario, as the packing fraction decreases until a critical value $\phi_c$, the system transitions from a jammed to an unjammed state at a jamming point $J$, whose quantitative value is known to depend on the driving protocol and boundary conditions. 

Another unifying framework for granular systems is the study of their rheological properties. Unjammed states have been effectively described by the inertial number $I$, a dimensionless parameter that relates a microscopic timescale of particle rearrangement to a macroscopic timescale associated with the deformation rate. Within this framework, the liquid-like state and its transition to a solid-like regime are described by a macroscopic effective interparticle friction coefficient $\mu$ and the packing fraction $\phi$, both depending on $I$~\cite{pouliquen2006flow-10e, man2023friction-dependent-4c9}.

 The complexity of the unjamming transition is illustrated by a pile of dry grains at rest. Even there, force distribution and mechanical equilibrium are nontrivial and strongly influenced by interparticle friction~\cite{papanikolaou2013isostaticity-bc5}. Interparticle friction allows the granular system to maintain stability over a wide range of packing fractions, with a lower bound known as random loose packing~\cite{onoda1990random-a16}. However, friction also introduces an indeterminacy in the contact network~\cite{roux2000geometric-bde}, and the system is only fully determined by knowing its preparation history. In addition, the critical number of contacts decreases with the static coefficient of interparticle friction~\cite{cruz2005rheophysics-b33}. 
Unlike fluids, stresses in granular media are highly heterogeneous, propagating along lines that form so-called force chains. These force chains, spanning several diameters, have been observed in numerous systems~\cite{peters2005characterization-3a5,vanel2000stresses-026}. However, the experimental observation of these chains in three-dimensional (3D) systems is considerably more challenging than in 2D~\cite{raynaud2002direct-298, philippe2002compaction-771}. 

In this context, we numerically investigate the internal behavior of a confined three-dimensional granular system near the unjamming transition. We use Discrete Element Method (DEM) simulations and a discrete and instantaneous particle extraction methodology. This extraction-driven protocol should be viewed as a controlled numerical thought experiment, designed to systematically approach marginal stability in a gravity-confined granular system.
For this procedure, in this article we provide: (i) a detailed characterization of the evolution of interparticle forces and system density as we extract particles; (ii) a novel quantification of the heterogeneity of contact forces by applying the Gini coefficient to a plasticity index~\cite{yitzhaki2012gini-b28,hurley2016quantifying-f96}, providing a metric to analyze the stress distribution; and (iii) a comparison between the unjamming transition and classical rheological descriptions of granular flows~\cite{pouliquen2006flow-10e}. This result indicates that, despite the fundamentally different driving mechanism and the focus on an extraction-driven unjamming process, the friction dependence of the critical packing fraction follows trends that are structurally consistent with those reported in rheological studies of dense granular flows.\\
By focusing on the approach to unjamming from the jammed side, our analysis highlights how interparticle friction shapes the loss of mechanical stability and the organization of force networks in gravity-confined granular systems, and enables meaningful qualitative comparisons with other gravity-driven and rheological configurations.\\
\section{\label{sec:numerProc}Numerical setup}
\paragraph{DEM simulations.} To investigate the mechanical response of a confined granular assembly subjected to particle removal, we perform three-dimensional Discrete Element Method (DEM) simulations using the open-source software \textit{MercuryDPM}~\cite{Weinhart2012,Weinhart2020}. In this framework, the translational and rotational motion of each particle is governed by Newton’s and Euler’s equations, respectively, while interparticle interactions are resolved through soft-sphere contact forces~\cite{cundall1979discrete-b93,Guo2015}. Gravity acts along the vertical ($z$) direction, introducing a preferred direction and an anisotropic stress state at the macroscopic scale.\\
Particle--particle and particle--wall contacts are modeled using a linear spring--dashpot formulation in both the normal and tangential directions~\cite{Luding2008,Shfer1996,Thornton2012}. In the normal direction, the contact force consists of an elastic contribution proportional to the particle overlap and a dissipative term proportional to the normal relative velocity. In this contact model, the elastic stiffness and the dissipative damping coefficient are directly related to the prescribed collision time and normal restitution coefficient. For a given effective mass, these two macroscopic collision parameters uniquely determine the constants appearing in the normal contact force law, ensuring consistent control of collision duration and energy dissipation.
Tangential forces follow an analogous spring--dashpot formulation and are constrained by a Coulomb friction criterion characterized by the static interparticle friction coefficient $\mu_p$.
To account for resistance to relative particle rotation---arising from surface roughness or slight deviations from perfect sphericity---we include a rolling friction model implemented in the MercuryDPM framework. This model introduces a restoring torque $M_r$ that opposes the relative rotation at the contact point and is limited by a Coulomb-like condition $|M_r| \le \mu_r R_{\mathrm{eff}} |f_n|,$
where $\mu_r = 0.05$ is the rolling friction coefficient, $R_{\mathrm{eff}}$ is the effective particle radius, and $f_n$ is the normal contact force. This mechanism provides an additional channel for dissipating rotational kinetic energy and contributes to the stabilization of the contact network.
The collision time is set to $t_{\mathrm{col}} = \SI{5e-4}{\s}$, and the integration time step is $\Delta t = \SI{1e-5}{\s}$, ensuring that collisions are well resolved and numerical stability is maintained. Both the normal and tangential restitution coefficients are fixed to $e = 0.5$, a value chosen to promote efficient dissipation of kinetic energy between successive particle extraction events. For the chosen values of the collision time $t_\text{col}$ and the restitution coefficient $e$, the normal spring stiffness and damping coefficient derived from the linear spring–dashpot model are $k=\SI{2032.23}{\newton/\m}$ and $\gamma = \SI{0.0136}{\newton \s/\m}$, respectively. Additional simulations performed with a larger restitution coefficient ($e = 0.8$), exhibit the same qualitative behavior, indicating that the results presented here are robust with respect to this parameter choice.

\paragraph{System preparation.}
The system consists of $N_0 = 30000$ spherical particles of diameter ${d = \SI{1}{\mm}}$ and material density $\rho = \SI{2500}{\kg\per\m\cubed}$, confined within a rectangular box with solid lateral walls of dimensions $L_x = L_y = 24d$. Particles are initially placed at random positions and allowed to settle under gravity until a mechanically stable configuration is reached. Using this preparation protocol, in which particles fall under gravity and settle into a mechanically stable configuration, the granular column reaches a height of approximately
$H_0\simeq 45d$. This height depends only weakly on the interparticle friction coefficient and on the particular realization. This state serves as the reference configuration from which a particle extraction process is initiated.
\begin{figure}[h!]
    \centering
    \includegraphics[width=0.5\linewidth]{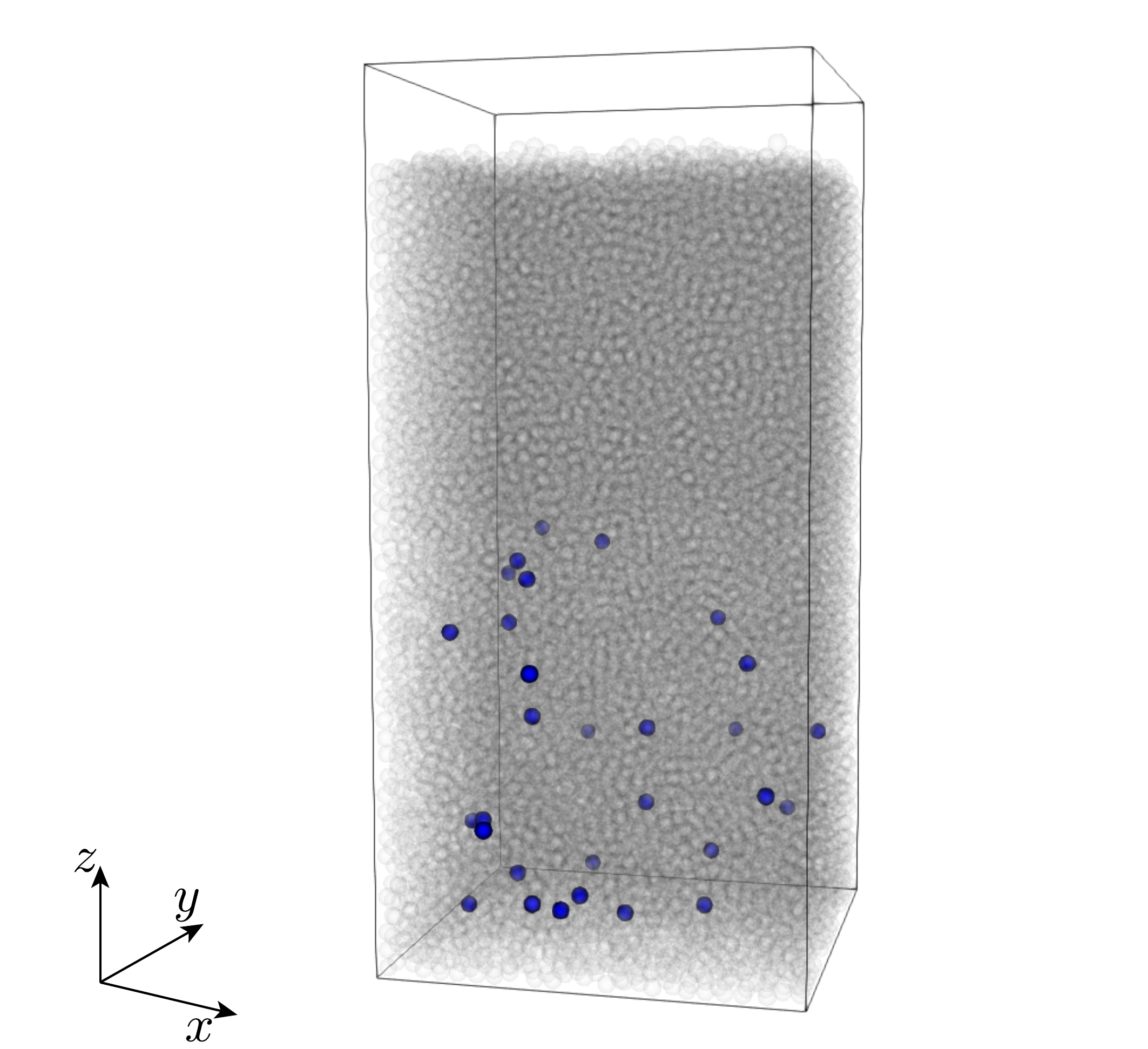}
    \caption{a) A snapshot of the numerical setup used in the simulations. A box with solid lateral walls whose size is $L_x= L_y=24d$ is initially filled with $N=30000$ spheres of diameter $d=\SI{1}{\mm}$. Every $\Delta t_e = \SI{0.02}{\s}$, $N_e = 30$ random particles are removed from the bottom half of the box (blue particles in the figure).}
    \label{fig:setup}
\end{figure}
\paragraph{Particle extraction protocol.}
Once the system has reached mechanical equilibrium, it is driven out of this initial state through a controlled particle-extraction procedure, conceptually similar to that employed by Rojas \textit{et al.}~\cite{rojas2023stability-7ff}. After the system has reached a mechanically stable configuration under gravity, at regular time intervals $\Delta t_e = \SI{0.02}{\s}$, $N_e = 30$ particles are randomly selected and instantaneously removed from the bottom half of the box. Removing particles exclusively from the lower region ensures that the reduction in packing fraction effectively propagates through the system via force-chain rearrangements, while avoiding direct perturbations of the free surface. An example of the selected set of particles at the moment of extraction prior to removal is shown in Figure ~1.
No external shear or directional forcing is applied. The evolution of the system is driven exclusively by gravity and the repeated, localized loss of particles. Simulations are performed for a total physical time of $\SI{4}{\s}$, which is sufficient to observe the full dynamical response induced by the extraction process. The static interparticle friction coefficient $\mu_p$ is used as a control parameter and is varied in the range ${\mu_p \in [0.2,\,0.9]}$, while all other parameters are kept fixed.\\
Thus, in our setup, the packing fraction is influenced by the particle-removal protocol. As particles are progressively removed, the total solid volume decreases while the volume of the box where the packing is computed, remains constant. We note that this particle-removal protocol is naturally suited to numerical simulations and is not straightforward to implement experimentally.\\

Unless otherwise stated, time and forces are expressed in dimensionless form by normalizing them with their characteristic values, $\sqrt{d/g}$ for time and $\rho d^3 g$ for forces.

\section{Results}\label{sec:results}
\subsection{Global response of the system and dynamical regimes}
To characterize the global response of the system to particle extraction, we monitor two macroscopic observables.
The first is the number of particles $N_K$ whose kinetic energy exceeds a fixed threshold defined as the amount of energy required to displace a particle by $1/5$ of its diameter, $ K>K_c \equiv \frac{1}{5}mgd$, 
where $m$ and $d$ denote the particle mass and diameter, respectively. This quantity provides a direct measure of the fraction of particles actively participating in the dynamics and serves as a sensitive indicator of motion triggered by the extraction events.
Second, we track the vertical position of the center of mass $z_{cm}$ of the granular column. This observable captures the global structural response of the system to particle removal and reflects how mass redistribution and internal rearrangements evolve over time.\\
\begin{figure}[t!]
    \centering
    \includegraphics[width=0.8\linewidth]{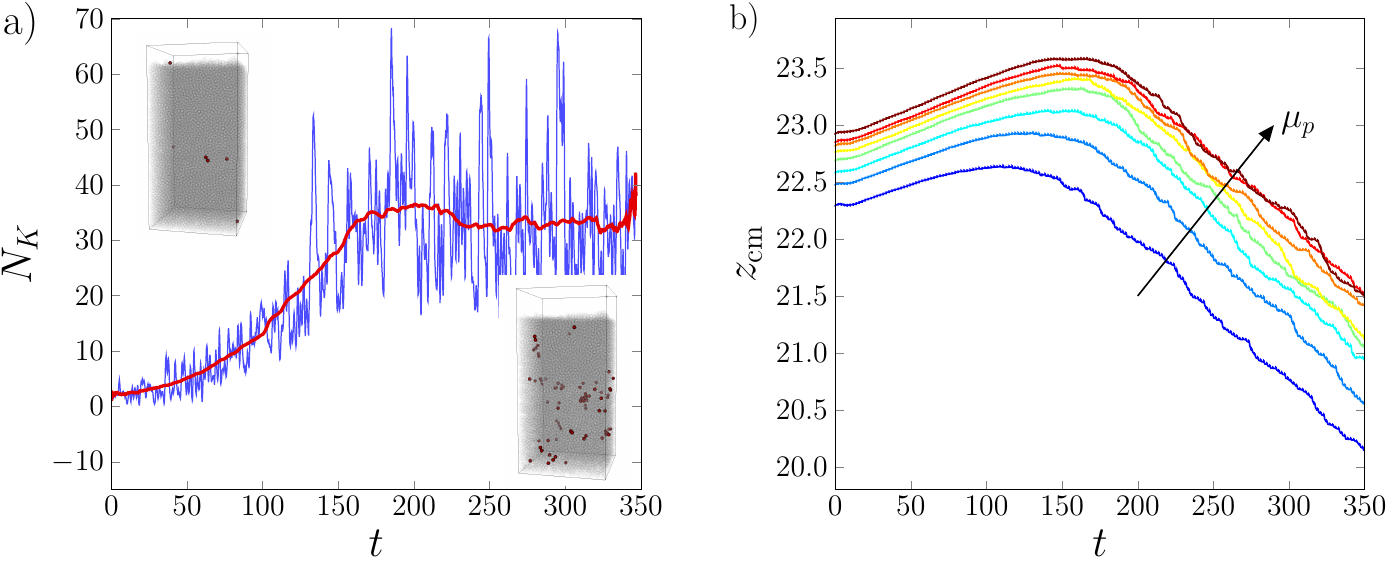}
    \caption{a)  Number of particles with kinetic energy exceeding the threshold 
$K_c = (1/5)\,mgd$ for $\mu_p = 0.8$. Two distinct regimes are observed during particle extraction: an initial regime in which this number increases in time, followed by a second regime where it reaches a saturation value. The red curve corresponds to the smoothed data. Insets show snapshots of the system in each regime; particles satisfying the kinetic-energy criterion are highlighted in red. b) Evolution of the $z-$coordinate of the center of mass (dimensionalized by the particle diameter $d$) over time for various friction coefficients. Each color corresponds to a different friction coefficient $\mu_p$ in the range ${0.2 \text{ (blue)}}$ to ${0.9\text{ (dark red)}}$.}
    \label{fig:setup2}
\end{figure}
As particles are progressively removed, the temporal evolution of $N_K$ and $z_{cm}$ (Figure~2a and 2b respectively) reveals the existence of \textit{two distinct dynamical regimes}. In an initial regime, particle motion remains limited: $N_K$ increases gradually with time, while 
$z_{cm}$
  displays a linear increase, since particles are extracted from the bottom half of the system, resulting in a progressive concentration of mass in the upper part. At later times, the system enters a second regime in which particle motion becomes sustained and spatially extended: $N_K$ reaches a steady value, and the center of mass decreases at an approximately constant rate which is independent of $\mu_p$. At this point, the system collapses every time particles are extracted, resulting in a downward motion that leads to a decrease in the vertical position of the center of mass.
Thus, from both panels of Figure~2, it can be inferred that at short times particle rearrangements are negligible, whereas at later times extraction events systematically trigger particle rearrangements throughout the system.
\\These two regimes are robustly observed across all friction coefficients explored and provide a global characterization of the system’s response to particle extraction.

\subsection{\label{sec:coordinationPacking}Coordination number and Packing fraction}

We now consider packing-related quantities: (i) the coordination number, which refers to the number of neighbors a particle has, and (ii) the packing fraction, referring to how efficiently particles occupy their available volume. 

\begin{figure}[b!]
    \centering
    \includegraphics[width=0.76\linewidth]{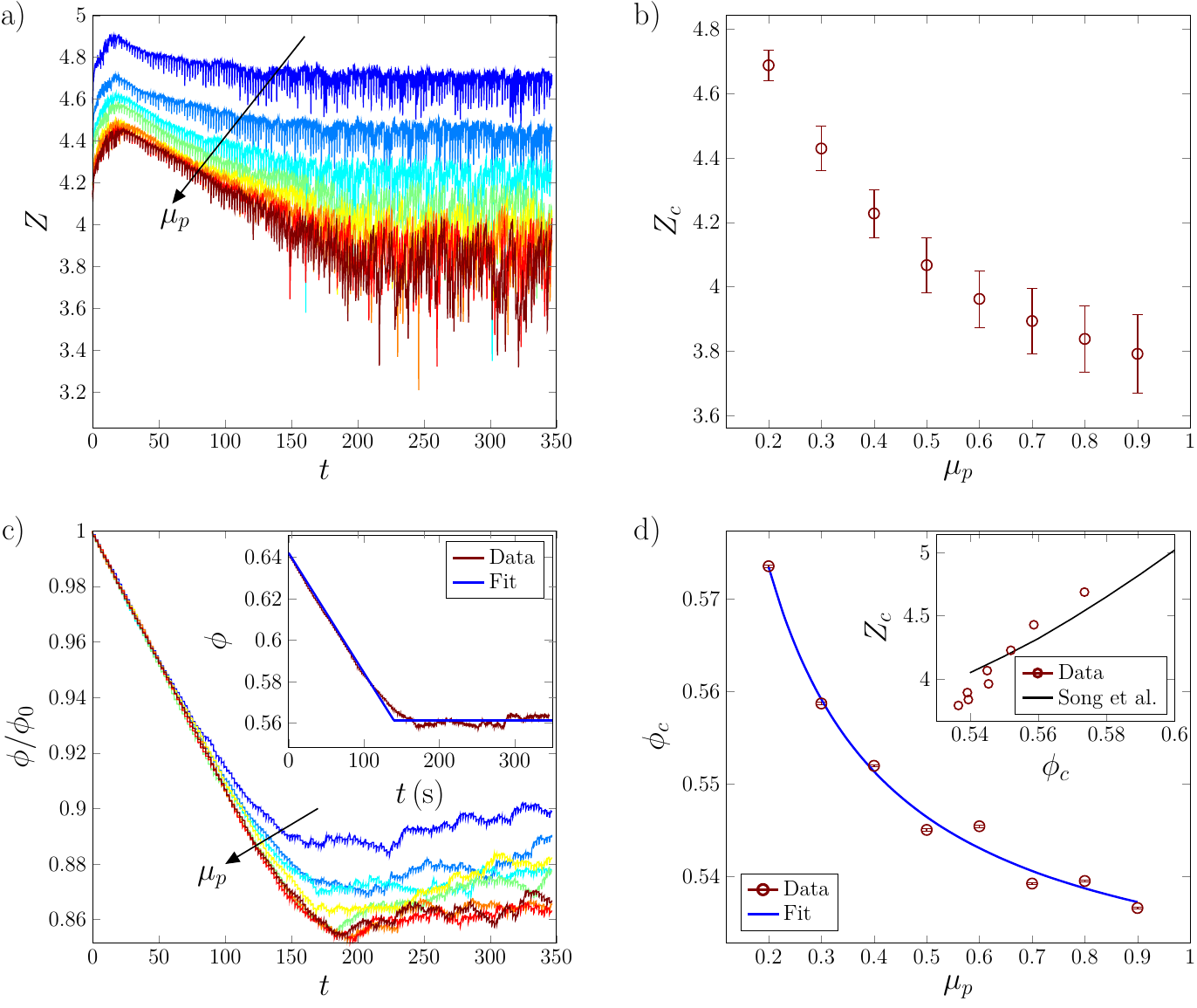}
    \caption{a) Mean coordination number versus time for different friction coefficients. The data show a monotonic, nearly linear decrease in the number of contacts as particles are removed, eventually reaching an asymptotic plateau $Z_c$. This critical value represents the mechanical limit of the system just prior to the unjamming transition. b) Critical coordination number as a function of interparticle friction. c) Packing fraction as a function of time. Inset shows and example of the fitting ${\phi(t) = \max(\phi_{0}(1-\alpha(t-t_c)),\,\phi_{c})}$ for $\mu_p=0.4$. We independently obtained $\alpha~=~9.5 \cdot 10^{-4}$ just by fitting a straight line for $t<60$. Each color corresponds to a different friction coefficient $\mu_p$ in the range ${0.2 \text{ (blue)}}$ to ${0.9\text{ (dark red)}}$. d) Critical packing fraction $\phi_c$ point as a function of the friction coefficient $\mu_p$. The fit corresponds to a rational function ${\phi_{c}=\frac{p_1\mu_p+p_2}{\mu_p+q_1}}$, with $p_1=0.5248$, $p_2=0.0329$, $q_1=0.0404$ and $R^2 = 0.9897$. The inset shows the critical coordination number $Z_c$ as a function of the critical packing fraction $\phi_c$. The solid line shown in the inset corresponds to the random loose packing prediction proposed by Song et al.~\cite{song2008phase-a4b}.}
    \label{fig:coord_number}
\end{figure}

As shown in Figure~\ref{fig:coord_number}a, the mean coordination number $Z$ exhibits a short initial transient, after which it decreases monotonically as particles are progressively removed from the system. This decay follows a nearly linear trend in the early stages of extraction, a behavior consistent with recent findings by Rojas et al. in a related granular system~\cite{rojas2023stability-7ff}. 
The transition from this linear decrease to a saturation plateau $Z_c$ is sensitive to the inter-particle friction; specifically, the onset of saturation occurs later for systems with larger friction coefficients. Furthermore, the critical value $Z_c$ decreases as the friction coefficient $\mu_p$ increases, as illustrated in Figure~\ref{fig:coord_number}b. This inverse relationship between $Z_c$ and $\mu_p$ is consistent with established results regarding the isostatic limits of frictional granular packings~\cite{silbert2010jamming-20a}, where higher friction allows for mechanical stability at a lower average number of contacts.

To study the evolution of the packing fraction $\phi$, we define a fixed imaginary cuboid of dimensions $20d \times 20d \times 23d$ horizontally centered in the lower region of the box. To ensure that our measurements are not influenced by boundary effects, the measurement volume is positioned at a distance of $2d$ from the lateral walls and the container base.
Initially, this volume contains $N^*_0 \approx 11100 $ particles (proportional to the cuboid-to-total volume ratio), then they decrease over time following a function $N^*(t)$. We compute $\phi$ as the ratio of the total volume occupied by the grains within this region to the total volume of the cuboid. By focusing on this localized internal volume rather than the entire container, we avoid the influence of the crystallized zones near the base and the density fluctuations at the top, thereby providing a representative value of the bulk packing fraction as the system approaches unjamming. Assuming that, during the initial stage of the extraction process, the particles within the cuboid stay fixed in their positions, the packing fraction decreases linearly with time following the rule
\begin{equation}
    \frac{\phi}{\phi_0} =\frac{N^*(t)}{N^*_0}
    =\frac{N^*_0-\frac{N^*_e}{\Delta t_e}t}{N^*_0}
\approx 1-9.55 \cdot 10^{-4}t,
\label{eq:phi_time}
\end{equation}
Here, $\phi_0$ denotes the initial packing fraction measured in the control volume after gravitational settling and prior to the onset of particle extraction. We note that $\phi_0$ depends weakly on the interparticle friction coefficient $\mu_p$, reflecting differences in the efficiency of compaction during the preparation stage. In Eq.~\ref{eq:phi_time} we have included the numerical values we used during the extraction process (initial number of particles $N^*_0$, the proportional number of extracted particles from the cuboid for each extraction ${N^*_e\approx21}$, and time step $\Delta t_e = 1.98$ for extractions). Figure~\ref{fig:coord_number}c shows that the initial decrease in $\phi$ is independent of the friction coefficient $\mu_p$, and occurs at a rate of $\alpha~=~9.5 \cdot 10^{-4}$, in very good agreement with the value found in equation \ref{eq:phi_time} obtained specifically for the present extraction protocol. At a later time, the system reaches a friction-dependent critical packing fraction $\phi_c$, after which the packing fraction fluctuates around a well-defined plateau despite continued particle removal. This behavior is observed consistently across all tested $\mu_p$ and is a signature of an unjamming transition~\cite{song2008phase-a4b}. Within this framework, unjamming is defined operationally as the onset of sustained particle rearrangements induced by successive extraction events; the corresponding critical values are therefore protocol dependent. By fitting the data with the function ${\phi(t) = \max(\phi_{0}(1-\alpha(t-t_c)),\,\phi_{c})}$, we obtain $t_c$ and $\phi_c$ (an example of the fit is shown in the inset of Figure~\ref{fig:coord_number}c). The time $t_c$ for reaching the plateau increases with $\mu_p$, while the critical packing fraction $\phi_c$ decreases as $\mu_p$ increases, as shown in Figure~\ref{fig:coord_number}d. The resulting values $t_c$ are summarized in Table~\ref{tab:unjamming_time}. We note that the critical packing fraction $\phi_c$ identified in our extraction-driven protocol is comparable to the solid concentrations at which Bagnold reported the onset of dense, inertia-dominated behavior in classical shear-flow experiments~\cite{bagnold1954experiments-a72}. Despite the very different driving mechanisms, both systems approach a loss of mechanical stability at similar structural densities.\\

\begin{table}[t]
\centering
\caption{Critical time $t_c$ for each interparticle friction coefficient $\mu_p$, obtained from fitting the packing fraction with ${\phi(t) = \max(\phi_{0}(1-\alpha(t-t_c)),\,\phi_{c})}$.}
\label{tab:unjamming_time}
\begin{tabular}{c c}
\hline
$\mu_p$ & $t_c$ \\
\hline
0.2 & $121.8$ \\
0.3 & $136.8$ \\
0.4 & $139.7$ \\
0.5 & $144.4$ \\
0.6 & $142.0$ \\
0.7 & $151.4$ \\
0.8 & $152.5$ \\
0.9 & $150.5$ \\
\hline
\end{tabular}
\end{table}

To summarize the results obtained so far, we observe two distinct regimes in the evolution of the packing fraction. At short times, when particle rearrangements remain limited, the removal of particles leads to a monotonic decrease of the packing fraction, consistent with a purely geometric effect. This regime coincides with the early-time behavior of the coordination number, which decreases approximately linearly as contacts are progressively lost. At later times, particle removal induces increasingly significant rearrangements throughout the system, as evidenced by the evolution of $z_{cm}$ and the number of dynamically active particles $N_K$ (Figure~2). In this regime, despite continued particle extraction, the packing fraction reaches a constant value, indicating that dynamical rearrangements compensate for the loss of solid volume, in close relation with the saturation observed in the coordination number.

Previous numerical and experimental studies of dense granular matter have shown that macroscopic properties near the onset of flow are strongly influenced by interparticle friction~\cite{cruz2005rheophysics-b33,song2008phase-a4b}. In particular, friction-dependent trends in both mechanical response and packing properties have been reported close to the transition between solid-like and flowing regimes. While the present work does not explicitly address steady granular flows, these results provide a useful reference for interpreting how microscopic friction affects critical properties at the threshold of mechanical stability.

Man et al.~\cite{man2023friction-dependent-4c9} numerically demonstrated a rational relationship between the microscopic particle friction coefficient $\mu_p$ and the corresponding macroscopic friction coefficient $\mu_1$, which for our purposes remains valid in the regime close to the transition. Building on this observation, we find that the critical packing fraction $\phi_c$ exhibits an analogous dependence on $\mu_p$, described by
\begin{equation}
\phi_c = \frac{p_1\mu_p + p_2}{\mu_p+q_1}.
\label{eq_phiMu}
\end{equation}
This trend reflects the intuitive notion that increasing interparticle friction reduces the system’s ability to pack efficiently, thereby lowering the density at which rigidity emerges. In this sense, the variation of the critical packing fraction with particle friction is consistent with trends reported within the $\mu(I)$ rheological framework originally proposed in~\cite{jop2006constitutive-4b6, pouliquen2006flow-10e}, despite the very different driving mechanism considered here. Finally, we verify that the robust critical point identified throughout this work represents the limit beyond which the system can no longer sustain its own stability, leading to a particle rearrangement once elements are removed. Intuitively, this point can be interpreted as the unjamming transition, corresponding to the random loose packing state of the system for a given friction coefficient. Moreover, $Z_c$ increases linearly with the critical packing fraction $\phi_c$ (see the inset in Figure~\ref{fig:coord_number}d), and aligns remarkably well with the random loose packing prediction proposed by Song et al.~\cite{song2008phase-a4b}, which is shown explicitly in the inset. Similar friction-dependent trends for $Z_c$ and $\phi_c$ have also been reported in studies of steady granular shear flows, despite the very different geometries and driving mechanisms involved~\cite{sun2011constitutive,berzi2015different,chialvo2012bridging}. While steady shear flows probe marginal stability under continuous deformation, the present extraction-driven protocol approaches the critical state from a dense, mechanically stable configuration under gravity. In this sense, our protocol provides an alternative route to marginal stability that leads to critical states consistent with those identified in shear-driven systems, within the expected dependence on geometry and driving conditions.

Once the system attains the value $\phi_c$, the following extractions trigger a vertical flow of grains to keep $\phi_c$ as a constant. This effect allows to predict the final slope of the vertical coordinate of the center of mass $z_{cm}$ (Figure \ref{fig:setup2}b). Indeed, if the packing fraction is constant, it is straightforward to verify from the definition of the packing fraction that:
\begin{equation}
    \frac{H_i}{N_i}=\text{const.},
\end{equation}
where $H_i$ denotes the total height of the granular column and $N_i$ represents the number of particles remaining in the system after the extraction $i$, respectively. The vertical coordinate of the center of mass is given by $z_{cm}^i = H_i/2$ for a homogeneous packing fraction, which allows to obtain an approximation of the decreasing rate of $z_{cm}$:
\begin{equation}
    \frac{\Delta z_{cm}}{\Delta t} \approx \frac{z_{cm}^0}{N_0}\frac{N_e}{\Delta t_e}.
    \label{Eq_dZ_dt}
\end{equation}
If we use equation \eqref{Eq_dZ_dt} to compute the decreasing rate of $z_{cm}$ for $\mu_p=0.2$, we obtain ${(\SI{22.3} {\mm} \cdot 30) /(30000 \cdot \SI{0.02}{\s}) = \SI{1.1} {\mm}/\SI{}{\s}}$ (or $0.011$ in its dimensionless form), in good agreement with Figure~\ref{fig:setup2}b. This outcome provides strong evidence that, under the extraction protocol used in the present study, the system ultimately sinks at a constant rate while maintaining a constant packing fraction.

Taken together, the results presented in this section provide consistent evidence that particle extraction drives the system from a mechanically stable (jammed) state toward a marginally stable, unjammed regime. This transition is characterized by the emergence of two clearly distinct dynamical regimes. At early times, particle removal produces only localized perturbations: the number of dynamically active particles $N_K$ increases slowly, the center of mass $z_{cm}$ rises linearly due to preferential extraction from the lower region, and both the packing fraction $\phi$ and the coordination number $Z$ decrease monotonically as a direct geometric consequence of particle loss. In this regime, particle rearrangements remain limited and do not compensate for extraction, indicating that the system largely preserves its load-bearing structure.

At later times, the system enters a qualitatively different regime in which particle extraction systematically triggers collective rearrangements. This regime is marked by a steady population of dynamically active particles, reflected by the saturation of $N_K$, and by a linear decrease of $z_{cm}$, indicating repeated downward reorganizations of the packing. Concomitantly, both the packing fraction and the coordination number reach well-defined, friction-dependent critical values, $\phi_c$ and $Z_c$, which remain constant despite continued particle removal. The saturation of these structural observables demonstrates that the system self-organizes into a marginal state in which rearrangements dynamically compensate for extraction, maintaining a constant density and connectivity.

Importantly, this late-time regime does not correspond to a fully flowing state: particle motion remains intermittent and spatially localized, and the system stays confined within the container. Nevertheless, the persistence of a finite fraction of active particles and the inability of the packing to sustain further loss of contacts or density without global rearrangement indicate that mechanical stability has been lost. In this sense, we identify the onset of this second regime as an extraction-driven unjamming transition, with $\phi_c$ and $Z_c$ representing the structural limits of stability for a given interparticle friction coefficient. Further evidence supporting this interpretation is provided in the following section through a direct analysis of force-chain statistics.

To elucidate the microscopic origin of this extraction-driven unjamming transition, we next analyze the evolution of interparticle forces and force-chain networks, which directly encode the loss of mechanical stability identified through the global observables discussed above.

\subsection{Forces}
In a pile of grains, the normal force amplitude spans a wide range of values~\cite{liu1995force-5bf}. A strong heterogeneity is observed between intense forces coexisting with small ones. Furthermore, the system's mechanical stability relies on two complementary sub-networks of contacts with a characteristic intensity in the forces~\cite{radjai1998bimodal-491}. Within these two sub-networks, there are normal forces where $f_n<\Bar{f}_n$ (soft network), and where $f_n>\Bar{f}_n$ (strong network), with $\bar{f}_n$ is the average normal inter-particle force in the system at a given time. In order to statistically characterize these forces, we compute the probability density function (PDF), also denoted by $P(\cdot)$. All the PDFs presented in this work are normalized such that the area under each curve is equal to unity.

\begin{figure}[b!]
    \centering
    \includegraphics[width=0.687\linewidth]{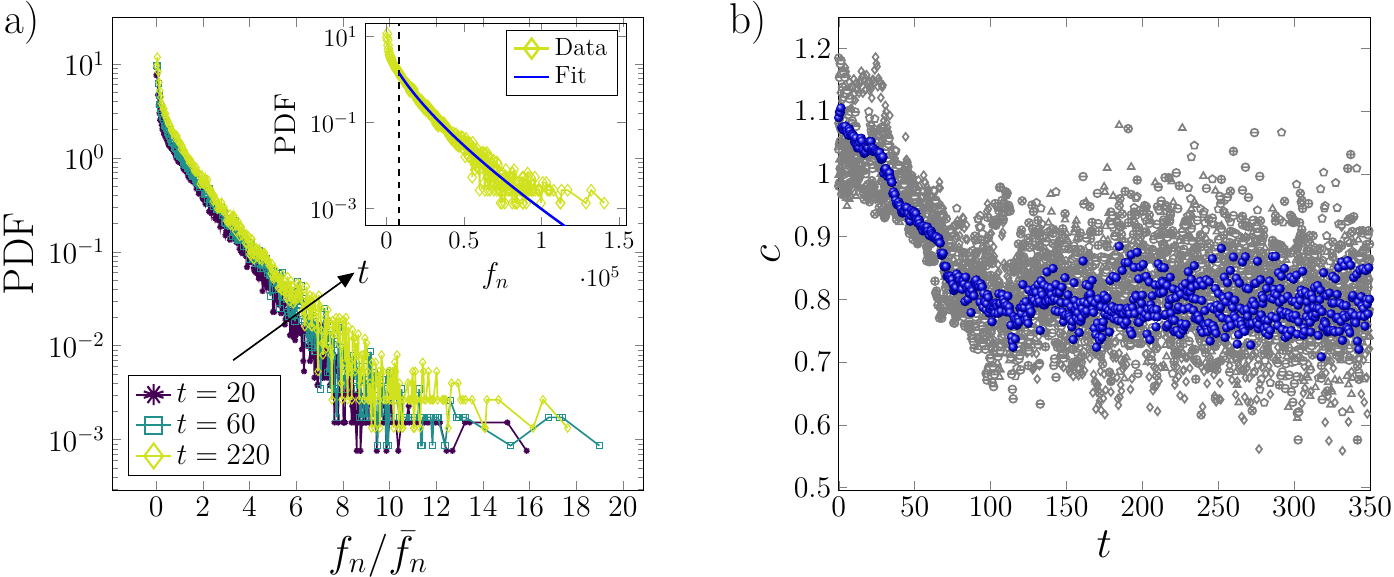}
    \caption{a) Probability Density Function (PDF) of the normal forces at three different times for $\mu_p=0.4$. All distributions are normalized to unit area ($\int P(\zeta) d\zeta = 1$). This distribution exhibits a power law for low forces and an exponential decay for high forces. The inset shows an example of the fit $\text{PDF}=A\exp(-(f_n/\bar{f}_n)^c)$. The dashed vertical line indicates the mean force for that particular data set. b) Time evolution of the exponent $c$. The data with blue markers represent the mean values averaged over all friction coefficients, while individual data sets for each $\mu_p$ are shown to illustrate the consistent independence of the particle friction coefficient. The value of $c$ quantifies the evolution of the force distribution, reflecting the dynamic reorganization of the contact network as particles are extracted.}
    \label{fig:pdf-exponent}
\end{figure}

The distribution of the normal contact forces, $P(f_n)$, provides insight into the internal structure of the load in the granular system. As seen in Figure~\ref{fig:pdf-exponent}a, $P(f_n)$ exhibits well-known characteristics: a possible power law for weak forces ($f_n < \bar{f_n}$), indicative of a large population of contacts that barely participate in the load distribution, and a broad tail for strong forces ($f_n > \bar{f_n}$). To characterize the tail evolution of this distribution as particles are extracted, we fit a stretched exponential function, $P(f_n)\sim \exp(-(f_n/ \bar{f_n} )^c)$, for $f_n>\bar{f_n}$ (see inset Figure~\ref{fig:pdf-exponent}a). 
Figure~\ref{fig:pdf-exponent}b shows the time evolution of the exponent $c$, which is found to be independent of the particle friction coefficient $\mu_p$. Initially, $c \approx 1$, representing a purely exponential decay characteristic of systems in equilibrium or dense states. As particles are extracted, $c$ decreases linearly until $t_c$, after which it reaches a plateau and stabilizes at $c = 0.8 \pm 0.1$. This transition signifies a robust structural reorganization of the contact network that is governed by the extraction process rather than surface friction. A $c<1$ indicates a ‘wider tail’ distribution (slower than exponential), suggesting a higher relative probability of encountering large forces compared to a purely exponential distribution. This change could be interpreted as a reorganization of the contact network where, as the number of contacts decreases (in agreement with~\cite{cruz2005rheophysics-b33}), the remaining forces become more heterogeneously concentrated to maintain system stability. The video included in the Supplementary Material illustrates the temporal evolution of the force network during particle extraction, showing how force-carrying structures progressively weaken and reorganize, with particles colored according to the magnitude of the force exerted on them (logarithmic scale).\\

\begin{figure}[t!]
    \centering
\includegraphics[width=1\linewidth]{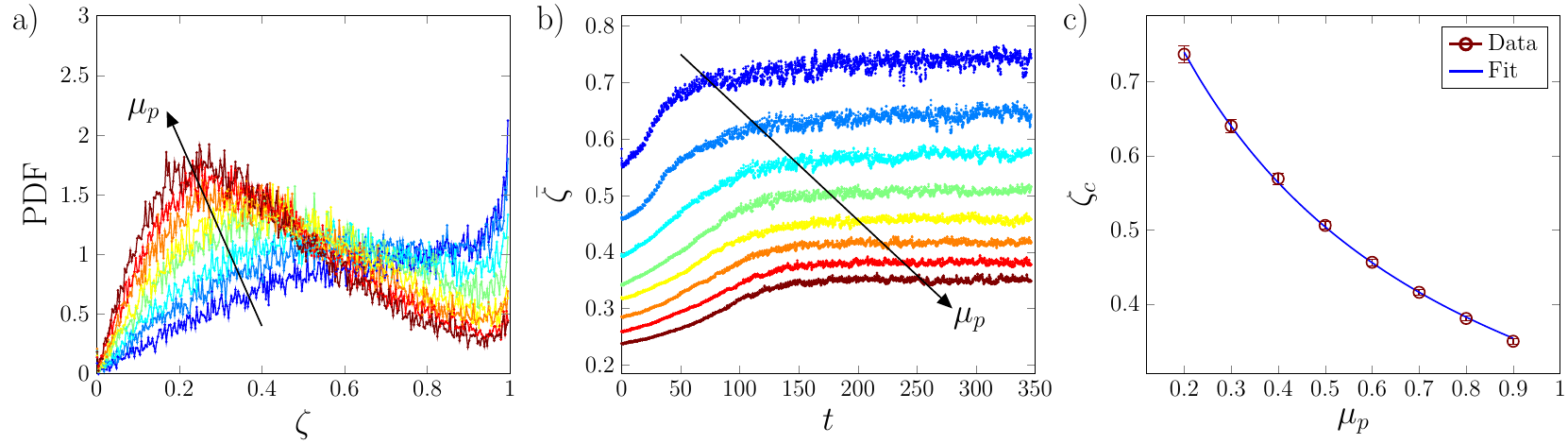}
    \caption{a) Probability density function of the plasticity index $\zeta\equiv f_t/\mu_p f_n$. The PDF are displayed at $t=300$. b) Average plasticity index $\langle \zeta \rangle \equiv \langle f_t/\mu_p f_n\rangle $ versus time. Each color corresponds to a different friction coefficient $\mu_p$ in the range ${0.2 \text{ (blue)}}$ to ${0.9\text{ (dark red)}}$.  c) Plasticity index once the system reached the critical point. The fit corresponds to a rational function $\zeta_c=\frac{p_0}{\mu_p+q_0}$, with $p_0=0.4747$, $q_0=0.4417$ and $R^2 = 0.9995$.}
    \label{fig:FactorCoulomb}
    \label{fig:pdf-cdf-coul}
\end{figure}

In Figure~\ref{fig:pdf-cdf-coul}a, we present the probability density functions (PDFs) of the plasticity index ${\zeta \equiv f_t/(\mu_p f_n)}$, where $f_t$ and $f_n$ are the tangential and normal components of the interparticle contact force, respectively \mbox{\cite{Silbert2002,silbert2010jamming-20a}}, evaluated at $t = 300$ (late-time regime). These distributions quantify the probability of finding a contact at a given state of frictional mobilization and are representative of the system’s behavior, as their shape is found to vary only weakly with time. Figure~\ref{fig:pdf-cdf-coul}a shows that all contacts satisfy $\zeta \leq 1$, consistent with the Coulomb non-slip condition. At low interparticle friction ($\mu_p = 0.2$), the distribution is relatively broad; however, it exhibits an upward tail as $\zeta \to 1$, indicating a significant population of contacts near the sliding threshold. As $\mu_p$ increases, the distribution develops a well-defined peak. The position of this maximum shifts toward lower values of $\zeta$ as $\mu_p$ increases, indicating that a higher interparticle friction coefficient allows the system to maintain mechanical stability with a larger proportion of contacts being far from the Coulomb failure limit.
We monitor the temporal evolution of $\zeta$ through its mean value $\bar{\zeta}$, which increases until reaching a steady-state plateau (see Figure~\ref{fig:pdf-cdf-coul}b). Consistent with the shift in the PDF peaks, $\bar{ \zeta }$ decreases as $\mu_p$ increases. The asymptotic values $\zeta_c$ are presented in Figure~\ref{fig:pdf-cdf-coul}c, confirming that the system reaches a final configuration dictated by the friction coefficient. This final state is reached at times in the range $\sim 80 -  140$ (depending on the interparticle friction coefficient $\mu_p$), coinciding with the stabilization of the normal force distribution exponents shown in Figure~\ref{fig:pdf-exponent}b.

\begin{figure}[b!]
    \centering
    \includegraphics[width=0.687\linewidth]{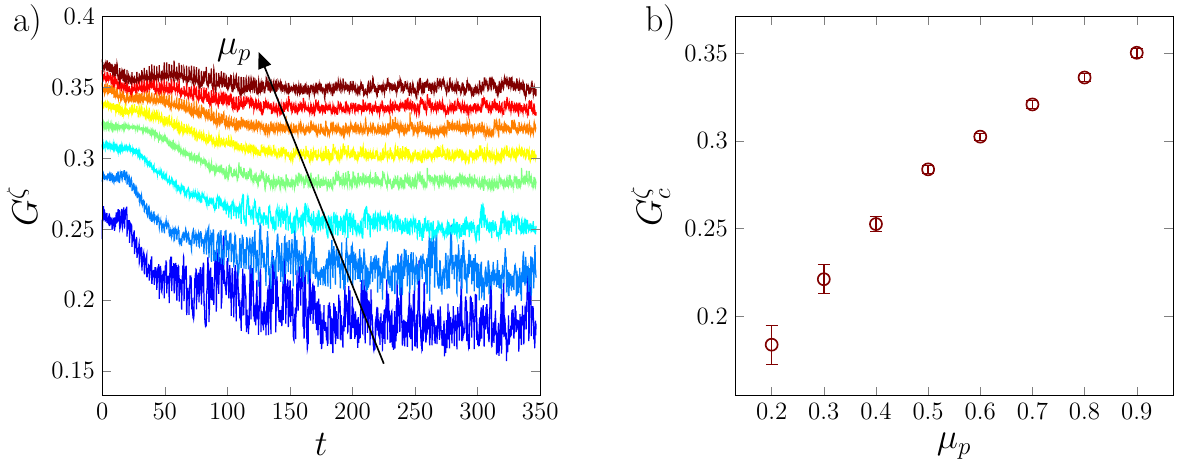}
    \caption{a) Gini factor of the plasticity index, $G^{\zeta}$ as a function of time for different friction coefficients $\mu_p$. Each color corresponds to a different friction coefficient $\mu_p$ in the range ${0.2 \text{ (blue)}}$ to ${0.9\text{ (dark red)}}$. b) Gini factor after the system reached the final jammed state $(t\gtrsim 100)$.}
    \label{fig:gini}
\end{figure}

To quantify the heterogeneity of the plasticity index $\zeta$ across the contact network, we compute its Gini coefficient $G^\zeta$. Originally developed to measure income inequality \cite{yitzhaki2012gini-b28}, this scalar metric has proven robust for characterizing the statistical dispersion of force distributions in granular media \cite{hurley2016quantifying-f96}. For a set of $n$ data points $\{x_1, x_2, \dots, x_n\}$ ordered increasingly ($x_1 \leq x_2 \leq \dots \leq x_n$), $G$ is defined as~\cite{gini1912variabilita, dixon1987bootstrapping}:

\begin{equation}
 G = \frac{1}{n} \left( n + 1 - \frac{2}{\sum_{i=1}^n x_i} \left[ \sum_{i=1}^n (n + 1 - i) x_i \right] \right)
\end{equation}

In the context of our system, the limits of $G^\zeta$ offer specific physical insights: 
\begin{itemize}
    \item \textbf{Low $G^\zeta$:} Indicates a \textit{homogeneous mobilization} where the ratio $\zeta$ is distributed evenly among contacts; the network shares the load collectively.
    \item \textbf{High $G^\zeta$:} Represents \textit{localized heterogeneity}, where a small subset of "strong" contacts carries the majority of the plastic load (approaching $\zeta \to 1$), while most other contacts remain poorly mobilized ($\zeta \approx 0$).
\end{itemize}
The Gini factor is particularly appropriate for this study because, unlike the standard deviation, it is non-dimensional and scale-independent, allowing for a direct comparison between simulations with different friction coefficients $\mu_p$ and varying contact numbers $n$. As shown in Figure \ref{fig:gini}, $G^\zeta$ decreases over time until reaching a critical value $G^\zeta_c$. This evolution indicates that the plasticity index becomes more homogeneous as particles are removed, while the higher $G^\zeta_c$ for larger $\mu_p$ confirms that interparticle friction facilitates the formation of more heterogeneous, chain-like force structures.
 
\subsection{\label{sec:force_chains}Force Chains}
\begin{figure}[b!]
    \centering
\includegraphics[width=0.687\textwidth]{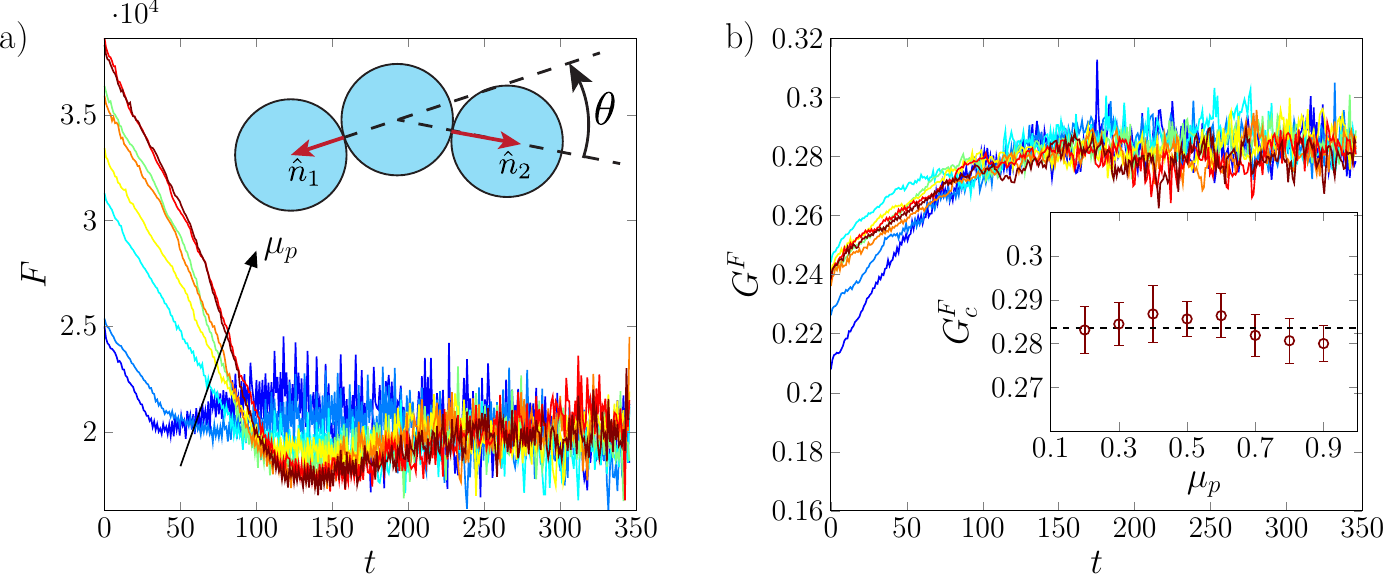}
    \caption{a) Chain forces intensity as a function of time for all the interparticle friction coefficients explored. b) Gini coefficient of the chain force intensity as a function of time. Inset shows the critical value of the Gini coefficient of the chain forces. The dashed line corresponds to the mean value of this coefficient ($\langle G^F_c\rangle = 0.284$) over all $\mu_p$. Each color corresponds to a different friction coefficient $\mu_p$ in the range ${0.2 \text{ (blue)}}$ to ${0.9\text{ (dark red)}}$.}
    \label{fig:chainintensity}
\end{figure}

Forces between grains in granular media are highly heterogeneous and tend to align spanning several particles' diameters~\cite{cruz2005rheophysics-b33, behringer2014statistical-20a, daniels2008force-fb7}. Aligned forces are called force chains. Among the various definitions of force chains, we use the one proposed by Peters et al.~\cite{peters2005characterization-3a5}: for a triad of particles $(i,j,k)$, we say they form a force chain if: (1) particles $i$, $j$ and $k$ are next neighbors; (2) the normal force $f_n$ between their contacts exceeds the mean value for all the particles (${f_n>\Bar{f}}$); (3) the angle $\theta$ is equal to or smaller than $\ang{30}$ (see inset Figure~\ref{fig:chainintensity}a). Using $\theta<\ang{30}$ in this definition, excludes branches in the force chains, making computation more straightforward.
As the system approaches the unjamming transition, the number of distinct force chains decreases in time (see Supplementary Material). This reduction indicates a loss of internal force-bearing capacity and a progressive breakdown of the contact network. A similar monotonic decrease in the number of force chains is observed when increasing interparticle friction coefficient $\mu_p$, suggesting that higher interparticle friction reduces the redundancy of force pathways needed to stabilize the packing. Quantitatively, the distribution of force chain lengths follows an exponential decay, consistent with prior observations in disordered granular media and indicative of short-range correlations in force transmission~\cite{peters2005characterization-3a5, zhang2014force-chain-803}.

Along with the decrease in force chain number, the average force magnitude within chains —i.e., the intensity $F$— diminishes as the system approaches the jamming threshold from above (see Figure~\ref{fig:chainintensity}a). In the jammed phase, the mean-force-chain intensity $F$ depends on $\mu_p$, reflecting the enhanced force anisotropy and inhomogeneity introduced by frictional interactions. However, this frictional dependence vanishes at the jamming point, suggesting a form of universality in the force network structure at the transition.
Importantly, as the system evolves from the jammed state toward unjamming, there is no evidence
of a drastic decrease of force chain length near unjamming (see Supplementary Material). Instead, the force network progressively weakens and fragments, with aligned force structures becoming increasingly short-lived. This behavior reflects the loss of mechanical stability under particle removal, rather than the emergence of long-range force correlations. The weakening and fragmentation of the force network as unjamming is approached has been reported in a variety of granular systems. Numerical and experimental studies have shown that, near unjamming, force chains lose temporal persistence and spatial extent, giving way to intermittent and highly heterogeneous stress transmission. For example, Radjai \textit{et al.}~\cite{radjai1998bimodal-491} and Behringer \textit{et al.}~\cite{behringer2014statistical-20a} emphasized that force networks close to marginal stability are characterized by fluctuating, short-lived force structures rather than stable chains. Similarly, simulations approaching the jamming threshold have shown that force-bearing networks progressively disintegrate as contacts are lost, without the development of diverging correlation lengths~\cite{silbert2010jamming-20a,OHern2003}. Our observations are consistent with this picture: when approaching unjamming from the jammed regime under particle removal, the system loses mechanical stability through fragmentation of the force network rather than through the growth of extended force chains.

We can access the evolving inhomogeneity of force transmission by computing the Gini coefficient of chain force intensities $G^{F}$, as we did for the plasticity index. Our simulations reveal that $G^{F}$ increases systematically as the system evolves toward the unjamming transition. This behavior indicates a progressive increase in the heterogeneity of stress distribution across the force chain network. Initially, in the jammed regime, the forces within the chains are evenly distributed, and many chains contribute comparably to the overall mechanical support. As the system approaches unjamming, the contact network becomes increasingly sparse, and many force chains weaken or entirely disappear.
Quantitatively, we observe that both the number of force chains and their mean intensity decrease as we approach the transition, consistent with the decline in normal forces at contact. This reduction leads to a growing disparity among the remaining chains: fewer chains sustain most internal stress, while most chains carry only weak forces or vanish altogether. This emergent inequality is captured by the increasing Gini coefficient, which reflects the redistribution of stress toward a small subset of dominant chains.
Despite varying the interparticle friction coefficient $\mu_p$ across a wide range in our simulations, the Gini coefficient exhibits a universal saturation value $G^{F}_c$ at the unjamming point: regardless of frictional strength, the system ultimately converges to a critical state characterized by maximal inhomogeneity in force transmission. The fact that this saturation value is independent of $\mu_p$ indicates that the final level of force heterogeneity is largely controlled by geometric constraints and the loss of mechanical stability, rather than by frictional dissipation. As shown in Figure~\ref{fig:chainintensity}b), the Gini factor of the force intensities converges to a common saturation value of approximately $0.284$ for all friction coefficients explored (see inset in Figure~\ref{fig:chainintensity}b).

\section{Discussion and Conclusions}\label{sec:concl}
This study advances the understanding of the unjamming transition in three-dimensional granular systems subjected to particle extraction by combining structural, statistical, and rheological perspectives. Using discrete element method (DEM) simulations, we characterized the evolution of interparticle forces, coordination number, and packing fraction throughout the extraction process. Both the packing fraction $\phi$ and the coordination number $Z$ systematically decrease, reaching critical values $\phi_c$ and $Z_c$ that depend on the interparticle friction coefficient $\mu_p$. These results highlight the role of the granular fabric and friction in guiding the system toward the structural instability within a given protocol, that marks the onset of unjamming.
To capture the heterogeneity of stress redistribution during this process, we introduced the Gini coefficient applied to the plasticity index $\zeta$. This measure evolves in time and stabilizes at a critical value $G^F_c \approx 0.28$ in the unjamming regime. The convergence toward this apparent universal-like value, independent of $\mu_p$, indicates that a small subset of contacts consistently bears most of the stress, reflecting the unequal load-sharing typical of force chain reorganization at the brink of instability.

Finally, the methodology employed here—progressive grain extraction—offers a novel route to determine the protocol-specific critical packing fraction $\phi_c$. Unlike the conventional approach, which estimates $\phi_c$ from the liquid state by fitting $\phi$ to the $\phi(I)$ rule in shear flows~\cite{cruz2005rheophysics-b33}, our procedure approaches the critical point from a dense solid phase. We emphasize that the critical packing fraction $\phi_c$ identified in this study is specific to the extraction-driven unjamming protocol considered here. As widely reported in the literature, the quantitative value of the unjamming density depends on the driving mechanism, boundary conditions, and preparation history~\cite{Silbert2002, Srivastava2022, lambert2025combined-910}. The relevance of our results lies in the fact that, within this protocol, $\phi_c$ is well defined and reproducible, and its dependence on interparticle friction follows trends consistent with those observed in other granular systems approaching marginal stability under different conditions.

Taken together, these results provide a coherent picture of extraction-driven unjamming as a protocol-dependent route to marginal stability, in which friction controls both the critical structural properties and the organization of force networks. Beyond their relevance for particle-removal processes, such as erosion or localized failure in granular assemblies, our findings contribute to a broader understanding of how unjamming emerges from the jammed state under non-shear-driven conditions.

\vskip6pt

\ack{We thank the supercomputing infrastructure of the High-Performance Computing UOH laboratory of Universidad de O'Higgins, Rancagua.}
\funding{Fondecyt Iniciación 11230970}.

\bibliographystyle{RS} 
\bibliography{references_final} 

@article{chialvo2012bridging,
  title={Bridging the rheology of granular flows in three regimes},
  author={Chialvo, Sebastian and Sun, Jin and Sundaresan, Sankaran},
  journal={Physical Review E—Statistical, Nonlinear, and Soft Matter Physics},
  volume={85},
  number={2},
  pages={021305},
  year={2012},
  publisher={APS}
}

@article{berzi2015different,
  title={Different singularities in the functions of extended kinetic theory at the origin of the yield stress in granular flows},
  author={Berzi, Diego and Vescovi, Dalila},
  journal={Physics of fluids},
  volume={27},
  number={1},
  year={2015},
  publisher={AIP Publishing}
}

@article{sun2011constitutive,
  title={A constitutive model with microstructure evolution for flow of rate-independent granular materials},
  author={Sun, JIN and Sundaresan, Sankaran},
  journal={Journal of Fluid Mechanics},
  volume={682},
  pages={590--616},
  year={2011},
  publisher={Cambridge University Press}
}

@article{lambert2025combined-910, 
  year     = {2025}, 
  title    = {Combined influence of particle friction and inertia on hysteresis in granular media on an inclined plane}, 
  author   = {Lambert, Clovis and Maurin, Raphaël and Lacaze, Laurent and Fede, Pascal}, 
  journal  = {Physical Review Fluids}, 
  doi      = {10.1103/physrevfluids.10.034301},
  pages    = {034301}, 
  number   = {3}, 
  volume   = {10}
}

@article{bagnold1954experiments-a72, 
  year    = {1954}, 
  title   = {Experiments on a gravity-free dispersion of large solid spheres in a Newtonian fluid under shear}, 
  author  = {Bagnold, Ralph Alger}, 
  journal = {Proceedings of the Royal Society of London. Series A. Mathematical and Physical Sciences}, 
  issn    = {0080-4630}, 
  doi     = {10.1098/rspa.1954.0186}, 
  pages   = {49--63}, 
  number  = {1160}, 
  volume  = {225}
}

@article{behringer2014statistical-20a, 
  year    = {2014}, 
  title   = {Statistical properties of granular materials near jamming}, 
  author  = {Behringer, R P and Bi, D and Chakraborty, B and Clark, A and Dijksman, J and Ren, J and Zhang, J}, 
  journal = {Journal of Statistical Mechanics: Theory and Experiment}, 
  doi     = {10.1088/1742-5468/2014/06/p06004}, 
  pages   = {P06004}, 
  number  = {6}, 
  volume  = {2014}
}

@article{Biroli2007, 
  year     = {2007}, 
  title    = {Jamming: A new kind of phase transition?}, 
  author   = {Biroli, Giulio}, 
  journal  = {Nature Physics}, 
  issn     = {1745-2473}, 
  doi      = {10.1038/nphys580}, 
  url      = {http://adsabs.harvard.edu/cgi-bin/nph-data_query?bibcode=2007NatPh...3..222B&link_type=ABSTRACT}, 
  pages    = {222 223}, 
  number   = {4}, 
  volume   = {3}, 
  language = {English}, 
  month    = {03}
}

@article{Castillo2012, 
  year     = {2012}, 
  title    = {Fluctuations and Criticality of a Granular Solid-Liquid-Like Phase Transition}, 
  author   = {Castillo, Gustavo and Mujica, Nicolás and Soto, Rodrigo}, 
  journal  = {Physical Review Letters}, 
  issn     = {0031-9007}, 
  doi      = {10.1103/physrevlett.109.095701}, 
  pmid     = {23002853}, 
  eprint   = {1204.0059}, 
  pages    = {095701}, 
  number   = {9}, 
  volume   = {109}, 
  language = {English}
}

@article{cruz2005rheophysics-b33, 
  year    = {2005}, 
  title   = {Rheophysics of dense granular materials: Discrete simulation of plane shear flows}, 
  author  = {da Cruz, Frédéric and Emam, Sacha and Prochnow, Michaël and Roux, Jean-Noël and Chevoir, François}, 
  journal = {Physical Review E}, 
  issn    = {1539-3755}, 
  doi     = {10.1103/physreve.72.021309}, 
  pmid    = {16196558}, 
  eprint  = {cond-mat/0503682}, 
  pages   = {021309}, 
  number  = {2}, 
  volume  = {72}
}

@article{cundall1979discrete-b93, 
  year    = {1979}, 
  title   = {A discrete numerical model for granular assemblies}, 
  author  = {Cundall, P A and Strack, O D L}, 
  journal = {Géotechnique}, 
  issn    = {0016-8505}, 
  doi     = {10.1680/geot.1979.29.1.47}, 
  pages   = {47--65}, 
  number  = {1}, 
  volume  = {29}
}

@article{daerr1999two-414, 
  year    = {1999}, 
  title   = {Two types of avalanche behaviour in granular media}, 
  author  = {Daerr, Adrian and Douady, Stéphane}, 
  journal = {Nature}, 
  issn    = {0028-0836}, 
  doi     = {10.1038/20392}, 
  pages   = {241--243}, 
  number  = {6733}, 
  volume  = {399}
}

@article{daniels2008force-fb7, 
  year    = {2008}, 
  title   = {Force chains in seismogenic faults visualized with photoelastic granular shear experiments}, 
  author  = {Daniels, Karen E. and Hayman, Nicholas W.}, 
  journal = {Journal of Geophysical Research: Solid Earth}, 
  issn    = {0148-0227}, 
  doi     = {10.1029/2008jb005781}, 
  number  = {B11}, 
  volume  = {113}
}

@article{dixon1987bootstrapping, 
  year    = {1987}, 
  title   = {Bootstrapping the Gini coefficient of inequality}, 
  author  = {Dixon, Philip M and Weiner, Jacob and Mitchell-Olds, Thomas and Woodley, Robert}, 
  journal = {Ecology}, 
  pages   = {1548--1551}, 
  number  = {5}, 
  volume  = {68}
}

@article{gans2023collapse-314, 
  year    = {2023}, 
  title   = {Collapse of a cohesive granular column}, 
  author  = {Gans, A. and Abramian, A. and Lagrée, P.-Y. and Gong, M. and Sauret, A. and Pouliquen, O. and Nicolas, M.}, 
  journal = {Journal of Fluid Mechanics}, 
  issn    = {0022-1120}, 
  doi     = {10.1017/jfm.2023.180}, 
  pages   = {A41}, 
  volume  = {959}
}

@book{gini1912variabilita, 
  title     = {Variabilità e mutabilità: contributo allo studio delle distribuzioni e delle relazioni statistiche}, 
  author    = {Gini, Corrado}, 
  publisher = {Tipogr. di P. Cuppini}, 
  address   = {Bologna}
}

@article{Guo2015, 
  year    = {2015}, 
  title   = {Discrete Element Method Simulations for Complex Granular Flows}, 
  author  = {Guo, Yu and Curtis, Jennifer Sinclair}, 
  journal = {Annual Review of Fluid Mechanics}, 
  issn    = {0066-4189}, 
  doi     = {10.1146/annurev-fluid-010814-014644}, 
  pages   = {1--26}, 
  number  = {1}, 
  volume  = {47}
}

@article{gutirrez2016oxford-a4b, 
  year   = {2016}, 
  title  = {Oxford Research Encyclopedia of Natural Hazard Science}, 
  author = {Gutiérrez, Francisco}, 
  doi    = {10.1093/acrefore/9780199389407.013.40}
}

@article{hurley2016quantifying-f96, 
  year    = {2016}, 
  title   = {Quantifying Interparticle Forces and Heterogeneity in 3D Granular Materials}, 
  author  = {Hurley, R. C. and Hall, S. A. and Andrade, J. E. and Wright, J.}, 
  journal = {Physical Review Letters}, 
  issn    = {0031-9007}, 
  doi     = {10.1103/physrevlett.117.098005}, 
  pmid    = {27610890}, 
  pages   = {098005}, 
  number  = {9}, 
  volume  = {117}
}

@article{Jaeger1996, 
  year    = {1996}, 
  title   = {Granular solids, liquids, and gases}, 
  author  = {Jaeger, {HM} and Nagel, {SR} and Behringer, {RP}}, 
  journal = {Reviews of Modern Physics}, 
  issn    = {0034-6861}, 
  doi     = {10.1103/revmodphys.68.1259}, 
  pages   = {1259 1273}, 
  number  = {4}, 
  volume  = {68}
}

@article{jop2006constitutive-4b6, 
  year    = {2006}, 
  title   = {A constitutive law for dense granular flows}, 
  author  = {Jop, Pierre and Forterre, Yoël and Pouliquen, Olivier}, 
  journal = {Nature}, 
  issn    = {0028-0836}, 
  doi     = {10.1038/nature04801}, 
  pmid    = {16760972}, 
  eprint  = {cond-mat/0612110}, 
  pages   = {727--730}, 
  number  = {7094}, 
  volume  = {441}
}

@article{Liu1998, 
  year    = {1998}, 
  title   = {Jamming is not just cool any more}, 
  author  = {Liu, Andrea and Nagel, Sidney}, 
  journal = {Nature}, 
  issn    = {0028-0836}, 
  doi     = {10.1038/23819}, 
  pages   = {1 2}, 
  number  = {6706}, 
  volume  = {396}, 
  month   = {11}
}

@article{liu1995force-5bf, 
  year    = {1995}, 
  title   = {Force Fluctuations in Bead Packs}, 
  author  = {Liu, C.-H. and Nagel, S. R. and Schecter, D. A. and Coppersmith, S. N. and Majumdar, S. and Narayan, O. and Witten, T. A.}, 
  journal = {Science}, 
  issn    = {0036-8075}, 
  doi     = {10.1126/science.269.5223.513}, 
  pmid    = {17842361}, 
  pages   = {513--515}, 
  number  = {5223}, 
  volume  = {269}
}

@article{Luding2008, 
  year    = {2008}, 
  title   = {Cohesive, frictional powders: contact models for tension}, 
  author  = {Luding, Stefan}, 
  journal = {Granular Matter}, 
  issn    = {1434-5021}, 
  doi     = {10.1007/s10035-008-0099-x}, 
  pages   = {235}, 
  number  = {4}, 
  volume  = {10}
}

@article{man2023friction-dependent-4c9, 
  year    = {2023}, 
  title   = {Friction-dependent rheology of dry granular systems}, 
  author  = {Man, Teng and Zhang, Pei and Ge, Zhuan and Galindo-Torres, Sergio A. and Hill, Kimberly M.}, 
  journal = {Acta Mechanica Sinica}, 
  issn    = {0567-7718}, 
  doi     = {10.1007/s10409-022-22191-x}, 
  pages   = {722191}, 
  number  = {1}, 
  volume  = {39}
}

@article{OHern2003, 
  year     = {2003}, 
  title    = {Jamming at zero temperature and zero applied stress: The epitome of disorder}, 
  author   = {O'Hern, Corey and Silbert, Leonardo and Liu, Andrea and Nagel, Sidney}, 
  journal  = {Physical Review E}, 
  issn     = {1539-3755}, 
  doi      = {10.1103/physreve.68.011306}, 
  pmid     = {12935136}, 
  eprint   = {cond-mat/0304421}, 
  url      = {http://adsabs.harvard.edu/cgi-bin/nph-data_query?bibcode=2003PhRvE..68a1306O&link_type=ABSTRACT}, 
  pages    = {11306}, 
  number   = {1}, 
  volume   = {68}, 
  language = {English}, 
  month    = {07}
}

@article{OHern2002, 
  year     = {2002}, 
  title    = {Random Packings of Frictionless Particles}, 
  author   = {O'Hern, Corey S and Langer, Stephen A and Liu, Andrea J and Nagel, Sidney R}, 
  journal  = {Physical Review Letters}, 
  issn     = {0031-9007}, 
  doi      = {10.1103/physrevlett.88.075507}, 
  pmid     = {11863912}, 
  eprint   = {cond-mat/0110644}, 
  url      = {http://adsabs.harvard.edu/cgi-bin/nph-data_query?bibcode=2002PhRvL..88g5507O&link_type=ABSTRACT}, 
  pages    = {75507}, 
  number   = {7}, 
  volume   = {88}, 
  language = {English}, 
  month    = {02}
}

@article{onoda1990random-a16, 
  year    = {1990}, 
  title   = {Random loose packings of uniform spheres and the dilatancy onset}, 
  author  = {Onoda, George Y. and Liniger, Eric G.}, 
  journal = {Physical Review Letters}, 
  issn    = {0031-9007}, 
  doi     = {10.1103/physrevlett.64.2727}, 
  pmid    = {10041794}, 
  pages   = {2727--2730}, 
  number  = {22}, 
  volume  = {64}
}

@article{papanikolaou2013isostaticity-bc5, 
  year    = {2013}, 
  title   = {Isostaticity at Frictional Jamming}, 
  author  = {Papanikolaou, Stefanos and O’Hern, Corey S. and Shattuck, Mark D.}, 
  journal = {Physical Review Letters}, 
  issn    = {0031-9007}, 
  doi     = {10.1103/physrevlett.110.198002}, 
  pmid    = {23705742}, 
  eprint  = {1207.6010}, 
  pages   = {198002}, 
  number  = {19}, 
  volume  = {110}
}

@article{peters2005characterization-3a5, 
  year    = {2005}, 
  title   = {Characterization of force chains in granular material}, 
  author  = {Peters, J. F. and Muthuswamy, M. and Wibowo, J. and Tordesillas, A.}, 
  journal = {Physical Review E}, 
  issn    = {1539-3755}, 
  doi     = {10.1103/physreve.72.041307}, 
  pmid    = {16383373}, 
  pages   = {041307}, 
  number  = {4}, 
  volume  = {72}
}

@article{philippe2002compaction-771, 
  year    = {2002}, 
  title   = {Compaction dynamics of a granular medium under vertical tapping}, 
  author  = {Philippe, P. and Bideau, D.}, 
  journal = {Europhysics Letters}, 
  issn    = {0295-5075}, 
  doi     = {10.1209/epl/i2002-00362-7}, 
  eprint  = {cond-mat/0210297}, 
  pages   = {677--683}, 
  number  = {5}, 
  volume  = {60}
}

@article{pouliquen2006flow-10e, 
  year    = {2006}, 
  title   = {Flow of dense granular material: towards simple constitutive laws}, 
  author  = {Pouliquen, O and Cassar, C and Jop, P and Forterre, Y and Nicolas, M}, 
  journal = {Journal of Statistical Mechanics: Theory and Experiment}, 
  doi     = {10.1088/1742-5468/2006/07/p07020}, 
  pages   = {P07020--P07020}, 
  number  = {07}, 
  volume  = {2006}
}

@article{radjai1998bimodal-491, 
  year    = {1998}, 
  title   = {Bimodal Character of Stress Transmission in Granular Packings}, 
  author  = {Radjai, Farhang and Wolf, Dietrich E and Jean, Michel and Moreau, Jean-Jacques}, 
  journal = {Physical Review Letters}, 
  issn    = {0031-9007}, 
  doi     = {10.1103/physrevlett.80.61}, 
  pages   = {61--64}, 
  number  = {1}, 
  volume  = {80}
}

@article{raynaud2002direct-298, 
  year    = {2002}, 
  title   = {Direct determination by nuclear magnetic resonance of the thixotropic and yielding behavior of suspensions}, 
  author  = {Raynaud, J. S. and Moucheront, P. and Baudez, J. C. and Bertrand, F. and Guilbaud, J. P. and Coussot, P.}, 
  journal = {Journal of Rheology}, 
  issn    = {0148-6055}, 
  doi     = {10.1122/1.1463420}, 
  pages   = {709--732}, 
  number  = {3}, 
  volume  = {46}
}

@article{rojas2023stability-7ff, 
  year    = {2023}, 
  title   = {Stability of a tilted granular monolayer: How many spheres can we pick before the collapse?}, 
  author  = {Rojas, Eduardo and Alarcón, Héctor and Salinas, Vicente and Castillo, Gustavo and Gutiérrez, Pablo}, 
  journal = {Physical Review E}, 
  issn    = {2470-0045}, 
  doi     = {10.1103/physreve.108.064904}, 
  pages   = {064904}, 
  number  = {6}, 
  volume  = {108}
}

@article{roux2000geometric-bde, 
  year    = {2000}, 
  title   = {Geometric origin of mechanical properties of granular materials}, 
  author  = {Roux, Jean-Noël}, 
  journal = {Physical Review E}, 
  issn    = {1539-3755}, 
  doi     = {10.1103/physreve.61.6802}, 
  pmid    = {11088375}, 
  eprint  = {cond-mat/0001246}, 
  pages   = {6802--6836}, 
  number  = {6}, 
  volume  = {61}
}

@article{Shfer1996, 
  year    = {1996}, 
  title   = {Force Schemes in Simulations of Granular Materials}, 
  author  = {Shäfer, J and Dippel, S and Wolf, D E}, 
  journal = {Journal de Physique I}, 
  issn    = {1155-4304}, 
  doi     = {10.1051/jp1:1996129}, 
  pages   = {5--20}, 
  number  = {1}, 
  volume  = {6}
}

@article{silbert2010jamming-20a, 
  year    = {2010}, 
  title   = {Jamming of frictional spheres and random loose packing}, 
  author  = {Silbert, Leonardo E.}, 
  journal = {Soft Matter}, 
  issn    = {1744-683X}, 
  doi     = {10.1039/c001973a}, 
  eprint  = {1108.0012}, 
  pages   = {2918--2924}, 
  number  = {13}, 
  volume  = {6}
}

@article{Silbert2002, 
  year    = {2002}, 
  title   = {Geometry of frictionless and frictional sphere packings}, 
  author  = {Silbert, Leonardo E. and Ertaş, Deniz and Grest, Gary S. and Halsey, Thomas C. and Levine, Dov}, 
  journal = {Physical Review E}, 
  issn    = {1539-3755}, 
  doi     = {10.1103/physreve.65.031304}, 
  pmid    = {11909043}, 
  pages   = {031304}, 
  number  = {3}, 
  volume  = {65}
}

@article{song2008phase-a4b, 
  year    = {2008}, 
  title   = {A phase diagram for jammed matter}, 
  author  = {Song, Chaoming and Wang, Ping and Makse, Hernán A.}, 
  journal = {Nature}, 
  issn    = {0028-0836}, 
  doi     = {10.1038/nature06981}, 
  pmid    = {18509438}, 
  pages   = {629--632}, 
  number  = {7195}, 
  volume  = {453}
}

@article{Srivastava2022, 
  year    = {2022}, 
  title   = {Flow and arrest in stressed granular materials}, 
  author  = {Srivastava, Ishan and Silbert, Leonardo E. and Lechman, Jeremy B. and Grest, Gary S.}, 
  journal = {Soft Matter}, 
  doi     = {10.1039/d1sm01344k}, 
  pages   = {735--743}, 
  volume  = {18}
}

@article{Thornton2012, 
  year    = {2012}, 
  title   = {Modeling of particle size segregation: Calibration using the discrete particle method}, 
  author  = {Thornton, Anthony and Weinhart, Thomas and Luding, Stefan and Bokhove, Onno}, 
  journal = {International Journal of Modern Physics C}, 
  issn    = {0129-1831}, 
  doi     = {10.1142/s0129183112400141}, 
  pages   = {1240014}, 
  number  = {08}, 
  volume  = {23}
}

@article{vanel2000stresses-026, 
  year    = {2000}, 
  title   = {Stresses in Silos: Comparison Between Theoretical Models and New Experiments}, 
  author  = {Vanel, L. and Claudin, Ph. and Bouchaud, J.-Ph. and Cates, M. E. and Clément, E. and Wittmer, J. P.}, 
  journal = {Physical Review Letters}, 
  issn    = {0031-9007}, 
  doi     = {10.1103/physrevlett.84.1439}, 
  pmid    = {11017537}, 
  eprint  = {cond-mat/9904094}, 
  pages   = {1439--1442}, 
  number  = {7}, 
  volume  = {84}
}

@article{Weinhart2020, 
  year    = {2020}, 
  title   = {Fast, flexible particle simulations — An introduction to {MercuryDPM}}, 
  author  = {Weinhart, Thomas and Orefice, Luca and Post, Mitchel and Lantman, Marnix P. van Schrojenstein and Denissen, Irana F.C. and Tunuguntla, Deepak R. and Tsang, J.M.F. and Cheng, Hongyang and Shaheen, Mohamad Yousef and Shi, Hao and Rapino, Paolo and Grannonio, Elena and Losacco, Nunzio and Barbosa, Joao and Jing, Lu and Naranjo, Juan E. Alvarez and Roy, Sudeshna and Otter, Wouter K. den and Thornton, Anthony R.}, 
  journal = {Computer Physics Communications}, 
  issn    = {0010-4655}, 
  doi     = {10.1016/j.cpc.2019.107129}, 
  pages   = {107129}, 
  volume  = {249}
}

@article{Weinhart2012, 
  year     = {2012}, 
  title    = {Closure relations for shallow granular flows from particle simulations}, 
  author   = {Weinhart, Thomas and Thornton, Anthony R and Luding, Stefan and Bokhove, Onno}, 
  journal  = {Granular Matter}, 
  issn     = {1434-5021}, 
  doi      = {10.1007/s10035-012-0355-y}, 
  pages    = {531 552}, 
  number   = {4}, 
  volume   = {14}, 
  language = {English}, 
  month    = {06}
}

@article{yitzhaki2012gini-b28, 
  year    = {2012}, 
  title   = {The Gini Methodology, A Primer on a Statistical Methodology}, 
  author  = {Yitzhaki, Shlomo and Schechtman, Edna}, 
  journal = {Springer Series in Statistics}, 
  issn    = {0172-7397}, 
  doi     = {10.1007/978-1-4614-4720-7}, 
  pages   = {11--31}
}

@article{zhang2014force-chain-803, 
  year    = {2014}, 
  title   = {Force-chain distributions in granular systems}, 
  author  = {Zhang, Ling and Wang, Yujie and Zhang, Jie}, 
  journal = {Physical Review E}, 
  issn    = {1539-3755}, 
  doi     = {10.1103/physreve.89.012203}, 
  pmid    = {24580218}, 
  pages   = {012203}, 
  number  = {1}, 
  volume  = {89}
}

\end{document}